\documentclass[amsmath,amssymb,twocolumn]{revtex4}
\usepackage{graphicx}
\usepackage{dcolumn}
\usepackage{bm}

\begin{document}

\title{X-ray pushing of a mechanical microswing}

\author{A. Siria$^{1,2}$, M. S. Rodrigues$^3$, O. Dhez$^3$, W. Schwartz$^{1,2}$, G. Torricelli$^4$, S. LeDenmat$^3$, N. Rochat$^{2}$, G. Auvert$^{2,5}$, O. Bikondoa$^3$, T. H. Metzger$^{3}$, D. Wermeille$^{3}$, R. Felici$^{3}$, F. Comin$^3$ and J. Chevrier$^{1}$}
\affiliation{$^{1}$ Institut N\'eel, CNRS-Universit\'e Joseph Fourier Grenoble, BP 166 38042 Grenoble Cedex 9, France\\
$^{2}$ CEA-LETI, 17 Avenue des Martyrs 38054 Grenoble Cedex 9, France\\
$^{3}$ ESRF, 6 rue Jules Horowitz 38043 Grenoble Cedex 9, France\\
$^{4}$ Department of Physics and Astronomy, University of Leicester, University Road Leicester LE1 7RH, England\\
$^{5}$ STMicroelectronics, 850 rue Jean Monnet, 38926 Crolles, France}

\date{\today}

\maketitle
\begin{figure}[t]
\includegraphics[width=1.\columnwidth]{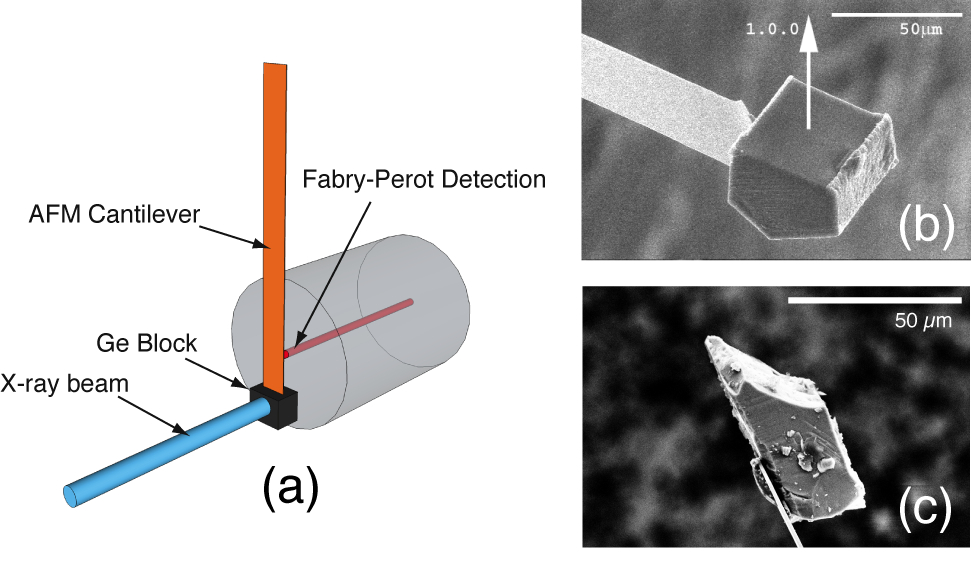} 
\caption{\label{fig:f1} {\bf Set-Up description}. (a) Schema of the experimental setup. Blue ray is the X-ray beam on the Ge micro-crystal at orange Si lever end. Grey cylinder represents the optical fiber and the red ray is the laser beam used to detect the lever position with sub-Angstrom precision.  (b and c) SEM image of the Ge cubes glued on Si levers. In (b) the cut and soldered Ge crystal using a Focus Ion Beam, has been positioned at the end of the lever in a symmetrical position (i.e. The COM Ge microcrystal is positioned below the lever end). In (c) a Ge crystal has been manually glued on the side in a very asymmetrical position.}
\end{figure}
\begin{figure}[t]
\includegraphics[width=1.\columnwidth]{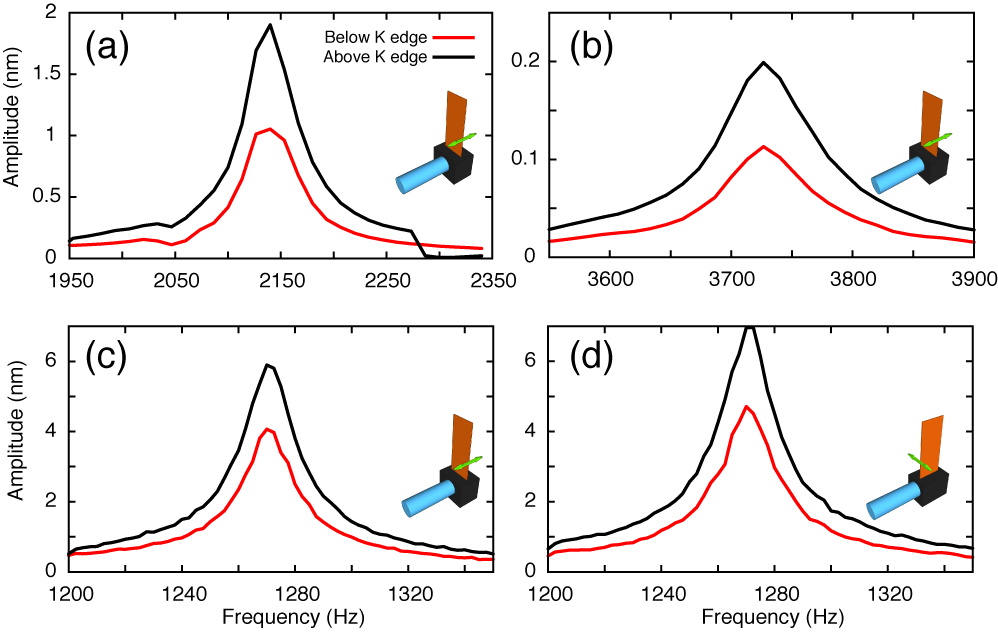}
\caption{\label{fig:f2} {\bf Measured resonance curve of the first oscillating mode for all levers}. In red the X-ray beam energy is set below the K1s edge ($E_{ph}=11.07 \: \mathrm{keV}$), in black it is set at the K1s edge ($E_{ph}=11.103 \: \mathrm{keV}$).\\
(a) Uncoated cantilever ($k=0.025 \: \mathrm{N}/ \mathrm{m}$, $Q=86$, $I_0=7.4 \, 10^{10} \: \mathrm{ph}/ \mathrm{s}$) with Ge block glued on the side and X-ray beam parallel to the oscillation direction.\\
(b) Coated cantilever ($k=0.027 \: \mathrm{N}/ \mathrm{m}$, $Q=60$, $I_0=3.5 \, 10^{10} \: \mathrm{ph}/ \mathrm{s}$) with Ge block glued on the side and X-ray beam parallel to the oscillation direction..\\
(c) Uncoated cantilever ($k=0.135 \: \mathrm{N}/ \mathrm{m}$, $Q=75$, $I_0=2.4 \, 10^{12} \: \mathrm{ph}/ \mathrm{s}$) with Ge block glued below and X-ray beam parallel to the oscillation direction.\\
(d) Same than (c) with X-ray beam perpendicular to the oscillation direction.}
\end{figure}

{\bf Nanoelectromechanical Systems (NEMS) are among the best candidates to measure interactions at nanoscale \cite{roukes06,ekinci04,yang06,Mamin01,Clealand98,rugar04}, especially when resonating oscillators are used with high quality factor \cite{burg07, craighead06}. Despite many efforts \cite{bargatin07,srini07}, efficient and easy actuation in NEMS remains an issue \cite{roukes01}. The mechanism that we propose, thermally mediated Center Of Mass (COM) displacements, represents a new actuation scheme for NEMS and MEMS. To demonstrate this scheme efficiency we show how mechanical nanodisplacements of a MEMS is triggered using modulated X-ray microbeams. The MEMS is a microswing constituted by a Ge microcrystal attached to a Si microcantilever. The interaction is mediated by the Ge absorption of the intensity modulated  X-ray microbeam impinging on the microcrystal. The small but finite thermal expansion of the Ge microcrystal is large enough to force a nanodisplacement of the Ge microcrystal COM glued on a Si microlever. The inverse mechanism can be envisaged: MEMS can be used to shape X-ray beams. A Si microlever can be a high frequency X-ray beam chopper for time studies in biology and chemistry.}\\
Previous studies of light mechanical effects on MEMS and NEMS have shown radiation pressure \cite{arcizet06} or thermal switch effect in the lever \cite{karrai04} as actuation mechanism for mechanical systems.  We show that these effects are not  effective enough to induce the observed oscillation amplitude in our experiments.\\
The experimental set-up is presented in fig. \ref{fig:f1}. The microswing position is measured through the interference between the light reflected from the back of the lever and from a cleaved fiber end. This experimental set-up has been shown to produce a sub-Angstrom precision in position measurements \cite{Clealand98,rugar04,karrai04}. SEM images  of the microswings used are shown in fig. \ref{fig:f1}(b) and \ref{fig:f1}(c).
Figure \ref{fig:f2} presents the mechanical response measured around the first resonance frequency $\omega_0$ for different geometries and experimental setups. The intensity of the X-ray beam impacting onto the Ge crystal is modulated at a  frequency $\omega$ sweeping through the lever resonant frequency $\omega_0$.  For X-ray energies below the absorption edge, the lever is already forced to oscillate with amplitudes larger than the thermally induced noise. 
For energies above the absorption edge we observe an increase of oscillation amplitude for all the geometries. The amount of this increase as function of geometry and microswing characteristic is the basis of our findings.

\begin{figure}[b]
\includegraphics[width=1.\columnwidth]{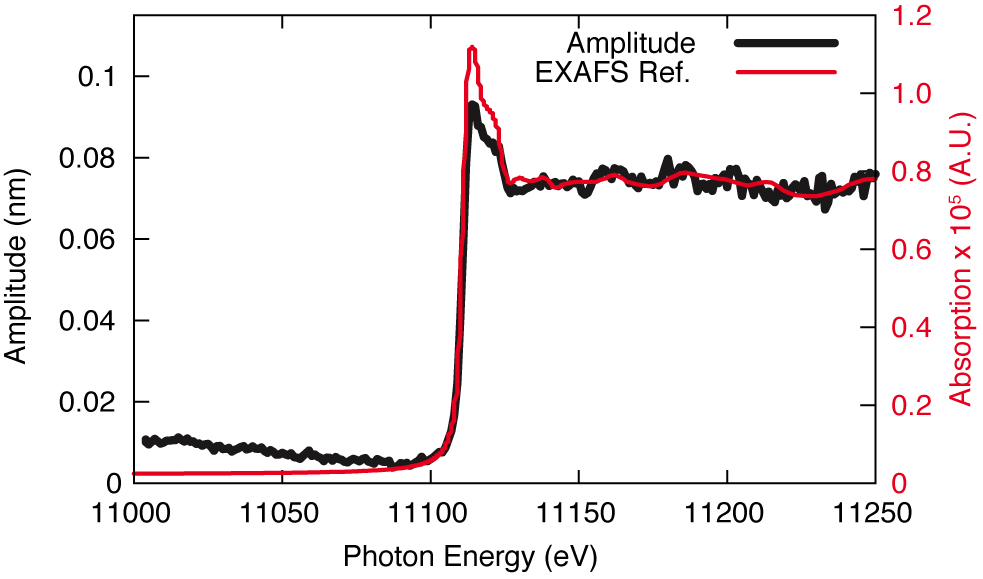}
\caption{\label{fig:f3} {\bf Cantilever oscillation amplitude in function of beam energy}. We show in black, our experimental data and in red, the handbook reference EXAFS spectrum at Ge K edge.}
\end{figure}

Figure \ref{fig:f3} reports the mechanical response of the cantilever at the resonance, when the X-ray energy is scanned through the germanium K-edge energy. The mechanical response of the microswing matches well the XAS reference spectrum of germanium \cite{nsls} . The two curves have been normalised below the edge and in the continuum atomic part above the edge. Even though a mechanical detection of EXAFS has already been shown \cite{masu89}, this is the first time utilising a MEMS. On the basis of the experimental evidence presented in fig. 2, we can identify the oscillation driving force. Radiation pressure can be ruled out as the oscillation is the same whatever the direction of the beam (figure \ref{fig:f2}(c) and \ref{fig:f2}(d)) with respect to the oscillation direction.\\
From fig. \ref{fig:f2} and \ref{fig:f3} it is evident that the oscillation amplitude is a function of the absorption cross section. Indeed, its spectrum follows well the absorption coefficient for germanium bulk. We explored then the hypothesis that the absorbed energy is promptly turned into heat leading to a temperature increase dependent on how the heat is evacuated.\\
As a first approximation, the number of photons that contributes to a temperature increase, is the difference between the number of absorbed photons and fluorescence photons that escape from the sample considering that the fluorescence emission can be photoelectrically reabsorbed. The overall number of photons $I_h$ that induce the temperature increase is then:
\begin{equation}
I_{h}=I_0 (1-T_{Ge}^{E})(1-w_{Ge}^{E} T_{w_{Ge}}^{E_f}) \label{eq:ih}
\end{equation}
where $I_0$ is the incoming intensity, $T_{Ge}^{E}$ the Ge transmission coefficient, function of the photon energy and sample thickness,  and $w_{Ge}^{E}$ the fluorescence yield. $T_{w_{Ge}}^{E_f}$ is the rate of fluorence at energy $E_f$ which escape from the sample. This last coefficient is dependent on sample thickness.\\
At energies below the Ge-K edge, the main process is the Auger electron production \cite{krause}. Most of the absorbed photons contribute then to the heating because of  short  mean free path (few nanometer) of the Auger electrons and their cascades. At energies higher than the Ge K edge the absorbed photons generate fluorescence, Coster-Kronig and Auger electrons.
\begin{table}[b]
\begin{ruledtabular}
\begin{tabular}{ccccccc}
\multicolumn{5}{c}{$l_0=23 \mu \mathrm{m}$}& Uncoated & Coated\\
$E_{ph}$&$T_{Ge}$&$w_{Ge}$&$T_{w_{Ge}}$&$I_h$&$x(\omega_0)\ [nm]$&$x(\omega_0)\ [nm]$\\ \hline
11.07 & 0.72 & 0 & - & 0.28 $I_0$& 1.053 & 0.113 \\
11.103 & 0.083 & 0.535 & 0.83 &0.51 $I_0$& 1.902 & 0.199\\
\multicolumn{3}{c}{}& \textbf{Ratio} & \textbf{1.82} & \textbf{1.81} & \textbf{1.76}\\
\multicolumn{7}{c}{}\\ \hline \hline
\multicolumn{5}{c}{$l_0=43 \mu \mathrm{m}$}& Paral. & Perp.\\
$E_{ph}$&$T_{Ge}$&$w_{Ge}$&$T_{w_{Ge}}$&$I_h$&$x(\omega_0)$&$x(\omega_0)$\\ \hline
11.07 & 0.54 & 0 & - & 0.47 $I_0 $& 4.066 & 4.713 \\
11.103 & 0.009 & 0.535 & 0.33 & 0.63 $I_0$ & 5.898 & 6.959\\
\multicolumn{3}{c}{}& \textbf{Ratio} & \textbf{1.34} & \textbf{1.47} & \textbf{1.48}\\
\end{tabular}
\end{ruledtabular}
\caption{{\bf Correspondance between absorbed photon and oscillation amplitude for different levers and geometries.} The top part presents the comparison , for a coated and an uncoated lever with an asymmetrical geometry like in Fig. 1(c). The X-ray beam is here parallel to the  direction of oscillation. The second part presents the comparison for an uncoated lever with a symmetric geometry. The X-ray beam is here either parallel or perpendicular to the direction of oscillation }
\label{tab:data}
\end{table}\\
The decreased amplitude of the XAFS peak and oscillations after the K-edge with respect to the reference spectra are due to this intrinsic self-absorption effect. In table \ref{tab:data} the absorbed photon flux $I_h$ is calculated for two lever/crystal configurations, for two X-ray beam directions, and for coated and uncoated levers. The ratio of the measured oscillation amplitudes $x(\omega_0)$ above and below K-edge energy is consistent with the ratio of absorbed photons.\\
The temperature increase $\Delta T$ can be calculated taking into account the overall energy deposited in the crystal and the heat flow throughthe lever (cooling by radiation and convection is here negligible). The absorbed power W is then:
\begin{eqnarray}
W=C\dot{T}(t)+G(T(t)-T_0) \label{eq:work}\\
T(t)=T_0+\frac{W}{G}\left(1-e^{-\frac{G}{C}t}\right) \label{eq:tem}
\end{eqnarray}
where $T_0$ is the ambient temperature and $T(t)$ the block temperature as function of time. $\Delta T(\omega)$ is then
\begin{eqnarray}
\Delta T(\omega) &=& \frac{W}{G} \frac{1}{(1+\omega\tau)} \label{eq:dt} \\
\tau&=&\frac{C}{G} \label{eq:tau1}
\end{eqnarray}
$\omega$ is the beam chopper frequency, $\tau$ is the ratio between the thermal capacity of the Ge block and the thermal conductivity of the Si lever. For the uncoated and the coated lever of (fig. \ref{fig:f2}(a) and \ref{fig:f2}(b)) the experimental conditions are nearly identical whereas the oscillation amplitude is 10 times larger in \ref{fig:f2}(a) than in \ref{fig:f2}(b). This difference can be described using those last equations. The presence of the metallic coating increases the thermal conductivity $G$ of the system and therefore induces a consequential decrease of  $\Delta T$ compared to the uncoated lever.\\
However this description cannot explain the difference of the amplitude of oscillation between the  (fig. \ref{fig:f2}(a) and \ref{fig:f2}(c)). The oscillation amplitude in fig. \ref{fig:f2}(c) is 3 times larger than in \ref{fig:f2}(c) against a photon flux 40 times bigger and an absorption rate 25\% higher because of the difference in Ge-crystal dimensions. The difference in the mechanical properties of the cantilever (\ref{fig:f2}(a) $k=0.025 \: \mathrm{N}/ \mathrm{m}$, \ref{fig:f2}(c) $k=0.135 \: \mathrm{N}/ \mathrm{m}$) cannot explain such a large descrepancy. However, the position of the Ge crystal and this symmetry with respect to the lever has not been considered. This remark is essential to  the conclusion of this paper. We show that the thermally induced change in the distance between the Ge crystal COM and the lever axis controls the system dynamics\\
The thermally induced change in the COM position is determined by :
\begin{eqnarray}
\Delta l(\omega)&=&l_0 \alpha \Delta T(\omega) \label{eq:dl}
\end{eqnarray}
$l_0$ is the distance between the block COM and the lever axis and $\alpha$ the linear thermal expansion coefficient.\\
For a simple 1D mechanical oscillator the oscillation amplitude is given by:
\begin{eqnarray}
x(\omega)&=&x_i(\omega)\sqrt{|\psi(\omega)|^2} \nonumber \\
&=&x_i(\omega)\sqrt{\frac{\omega^2_0 Q^2}{\frac{Q^2}{\omega^2_0}(\omega^2-\omega^2_0)^2+\omega^2}} \label{eq:x2}
\end{eqnarray}
$\psi(\omega)$ is the oscillator transfer function. Here, $x_i(\omega)$  corresponds to $\Delta l(\omega)$.\\
\begin{figure}[b]
\includegraphics[width=1.\columnwidth]{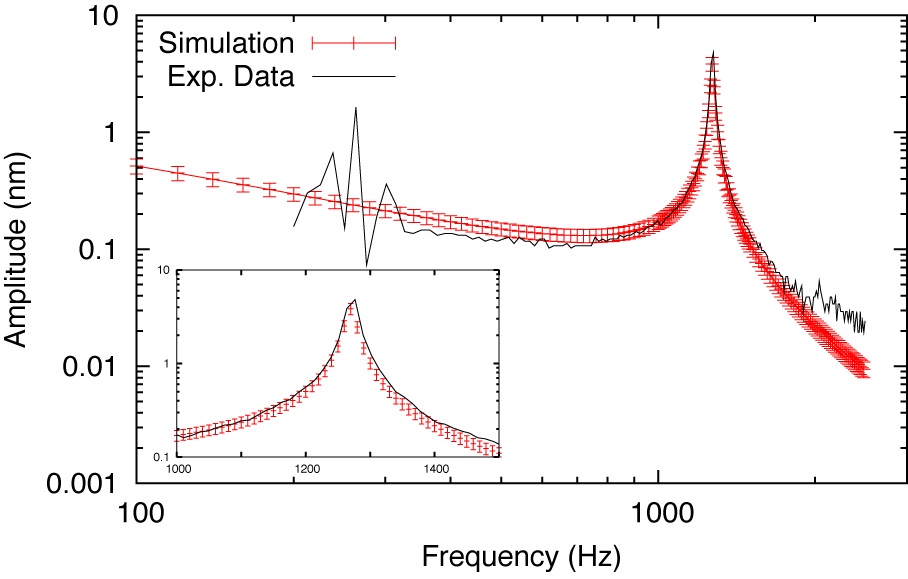}
\caption{\label{fig:sim} {\bf Response function of the lever shown in figure 1(b)}. Black curve is the measured amplitude of the lever oscillation as the beam intensity is modulated from 100 Hz to 2500 Hz . Red curve is the calculated expression using experimental parameters characteristics of the X-ray beam, of the X-ray absorption around the Ge K edge and of the lever described as a single mode oscillator. The error bar in red curve has been determined using the Brownian motion. Red curve calculation involves the misalignement of the Ge microcrystal on the Si lever as the single adjustable parameter. In the inset a zoom on the resonant peak is presented.}
\end{figure}
For the system in fig. \ref{fig:f1}(c), $l_0=13 \mu m$ close to half the Ge crystal thickness. For an  intensity $I_0=7.4\ 10^{10} \: \mathrm{ph}/\mathrm{s}$ the  temperature increase is found to be $\Delta T(\omega_0)=0.24 \: \mathrm{K}$. Using $\alpha_{Ge}=5.9 \, 10^{-6} \: \mathrm{K}^{-1}$, according to equation \ref{eq:dl}, the induced COM displacement is $\Delta l(\omega_0)=19 \: \mathrm{pm}$. Using equation \ref{eq:x2}, with the measured quality factor of 86 and the amplitude at the resonance of $1.9 \: \mathrm{nm}$, the COM displacement is found to be $\Delta l(\omega_0)= 22 \: \mathrm{pm}$ which is consistent with the value calculated from equation \ref{eq:dl}. The error bar on the measured lever position is determined by the thermal fluctuations of the lever position and is $x_i(k_B T) = 1.6 \: \mathrm{pm} $.\\
The system in fig. \ref{fig:f1}(b) presents a much more symmetrical geometry.  $l_0$ value in this case must be smaller than the one in the case of fig. \ref{fig:f1}(c), but it is not easily measurable. A rough estimate of the residual misalignment between the COM of Ge microcrystal and the Si lever axis is the incertitude in the FIB positioning device that is about 1 $\mu m$.\\
The distance $l_0$ that best fits the data while all other parameters are known is 1.5 $\mu m$ which is indeed close to the precision of the FIB motor. The comparison between the model (equation \ref{eq:x2}) and the measured oscillation is presented in figure \ref{fig:sim} as the excitation frequency is swept from 100 Hz to 2500 Hz. The agreement further establishes that the thermally forced displacement of the COM is at the origin of the observed lever oscillation equiped with the Ge crystal. Results for all configurations are then consistently explained using this single actuation mechanism.\\
The MEMS actuation mechanism shown here can be extended to NEMS actuation. Considering a Si lever of $1\times0.1\times0.1$ $\mu$m and a Ge block of $100\times100\times100$ nm with thermal conductivity of $G=3.7 \cdot 10^{-8} \: \mathrm{W}/\mathrm{K}$ and thermal capacity of $C=1.7 \cdot 10^{-15} \: \mathrm{J}/\mathrm{K}$ \cite{Li03} leads, according to Eq. \ref{eq:dt}, to a substantial temperature increase at a frequency in the MHz regime, typical for the
resonance of such a NEMS. If a 1 $\mu W$ laser beam is absorbed in this Ge block, the induced thermal expansion will be several pm. As NEMS with lateral size close to $100\: \mathrm{nm}$ can exhibit quality factors of 1000, a forced COM oscillation with amplitude of several pm can result at resonance in a nanometric NEMS oscillation amplitude. This is far above the thermally induced fluctuations of NEMS position. This strategy of NEMS excitation can be compared to photothermal actuation based on thermally induced strain \cite{koenig06}. The essential difference is in the origin of the NEMS displacement. This origin is, in the mechanism that  we propose, a strain-free thermally induced change in mass spatial distribution in asymmetric structure.\\
Due to limited performances of current X-ray choppers, MEMS is here operated close to 1 $\mathrm{kHz}$ i.e. at very low frequencies. The use of MEMS as Si single crystal micro-oscillators can provide X-ray choppers at much higher frequencies. We have already produced experimental evidence of such an effect at 13 $\mathrm{kHz}$. Using diffraction, Si single crystal MEMS appear as a good candidate for the high frequency manipulation of X-ray microbeam. This could offer new tools either to change phase X-ray wavefront, or to produce rapidly modulated intensity of X-ray beams that are so important in real time studies of fast dynamical processes in chemistry and in biology \cite{wulff97}. 
\section{METHODS}
\subsection{MICROSWING REALISATION}
The first Ge microcrystal in fig. \ref{fig:f1}(b) has been directly cut to Ge wafer by a Focus Ion Beam (FIB). In order to fabricate the micro-oscillator, a cubic like germanium crystal has been etched from a (1 0 0) oriented germanium wafer using the FIB Strata400 from FEI. Then, the cube has been extracted using a motorized tungsten tip and placed close to the silicon cantilever end. Finally, the cube was welded to the cantilever using localized FIB deposition of metal. The cubic Ge crystal is 43 $\mu m$ thick. It is soldered to the Si lever in a symmetrical position. The lever is a standard Silicon AFM cantilever whose dimensions are 350x35x2 $\mu m^3$. This lever has no metallic coating. The second Ge microcrystal is about 23 $\mu m$ thick (fig. \ref{fig:f1}(c)). It has been manually glued on the side of the cantilever in a very asymmetrical position. For asymmetrically mounted crystals, two types of levers have been used: one bare and another with a metallic coating.
\subsection{BEAMLINE SET-UP}
The experiences were performed at the European Synchrotron Radiation Facility (ESRF). The beamlines involved were the Anomalous Scattering Beamline (ID01) and Surface Science X-Ray Diffraction (SXRD) Beamline (ID03). The Anomalous Scattering Beamline ID01 has been designed to combine small and wide angle X-ray scattering techniques with anomalous diffraction. The radiation from the undulators can be tuned from 2.5 to 40 keV with a Si(111) double crystal monochromator (energy resolution $\Delta E/E\approx10^{-4}$). Focusing is achieved by using beryllium Compound Refractive Lenses (CRLs) \cite{snigirev96}. The effective focus size is $\approx 4\times 6 \: \mu m^2$ with $\approx 10^{10}$ photons/second on the focal spot. At the SXRD beamline the photons were tuned at the Ge K edge using a liquid nitrogen cooled monolithic double crystal Si (111) monochromator. The beam was focused at the sample by a Kirkpatrick-Baez (KB) mirror system located 43 m from the photon source. The beam size at the sample, 1 m from the KB system, is $\approx 3\times 5 \: \mu m^2$ with $\approx 10^{12}$ photons/second on the focal spot.

\end{document}